# Characteristics of Gravity Waves in Opposing Phases of the QBO: A Reanalysis Perspective with ERA5

Hamid A. Pahlavan,[a,b] John M. Wallace,[a] Qiang Fu,[a] M. Joan Alexander [c]

[a] *University of Washington, Seattle, Washington*

[b] *Rice University, Houston, Texas*

[c] *NorthWest Research Associates, Boulder, Colorado*

*Corresponding author*: Hamid A. Pahlavan, pahlavan@rice.edu

ABSTRACT

ERA5 data for the period of 1979-2019 are used as a basis for investigating the properties of gravity waves as they disperse and propagate upward through the stratosphere during opposing phases of the QBO. Two-sided zonal wavenumber-frequency spectra of vertical velocity in the stratosphere exhibit distinctive gravity wave signatures. Consistent with theory, westward propagating waves tend to be suppressed during the easterly QBO phase and eastward propagating waves tend to be suppressed during the westerly phase. Cospectra of the vertical flux of zonal momentum also show significant asymmetries between eastward and westward propagating waves during opposing QBO phases. Phase speed spectra of the vertical flux of momentum are indicative of a strong dissipation of westward propagating gravity waves during the easterly phase and vice versa; i.e., a selective "wind filtering" of the waves as they propagate upward. The three-dimensional structure of the gravity waves is revealed by compositing. In the absence of a background zonal flow, the waves radiate outward and upward from their respective reference grid points in concentric rings. When a zonal flow is present, the rings are amplified and compressed upstream of the source and attenuated and stretched downstream of it, such that they assume the form of arcs. These results serve to confirm the applicability of the mechanism proposed by Lindzen and Holton (1968) to explain the downward propagation of the QBO. The QBO also influences the spectrum of gravity waves in ERA5 at 100 hPa, below the layer in which wind-filtering occurs.

## 1. Introduction

Tropical convection excites a broad range of waves, including high-frequency, small-scale gravity waves that impact the global atmosphere by transporting energy and momentum over large horizontal and vertical distances, and act as an important coupling mechanism between atmospheric layers (Fritts and Alexander 2003). Small-scale gravity waves contribute to the wave-driving of the tropical stratospheric quasi-biennial oscillation (QBO), particularly during the descent of the easterly shear zones (Sato and Dunkerton 1997; Piani et al. 2000; Giorgetta et al. 2002; Kawatani et al. 2010; Alexander and Ortland 2010; Holt et al. 2020; Pahlavan et al. 2021a,b). However, the limited vertical resolution of the observations and the uncertainties in the representation of gravity waves in numerical models limit our understanding of their sources, propagation, and dissipation.

In the companion paper (Pahlavan et al. 2023) we showed that the ERA5 Reanalysis (Hersbach et al. 2020), with a horizontal resolution of 0.28° (~31 km), 137 hybrid model levels, and hourly temporal resolution, resolves a larger fraction of the gravity wave spectrum than its predecessors and thus provides valuable information on their behavior and spectral characteristics. In this paper, we extend the previous study by contrasting the properties of gravity waves in the tropical stratosphere during opposing phases of the QBO during the period of 1979-2019, which spans ~16 QBO cycles. According to the mechanism proposed by Lindzen and Holton (1968) to explain the downward phase propagation of the QBO, we expect that eastward propagating waves will be absorbed preferentially in westerly shear zones of the QBO and westward propagating waves in easterly shear zones.

Several previous studies have reported correlations between the QBO phase and gravity wave activity based on station data. Sato and Dunkerton (1997) estimated momentum fluxes by gravity waves based on an 8-year time series of radiosonde observations at Singapore. Vincent and Alexander (2000) carried out a similar study of small-scale waves based on a 6-year data set of radiosonde observations at Cocos Islands. In addition to strong annual and interannual variations they found OBO-related variations in wave energy and zonal momentum fluxes.

de la Torre et al. (2006) and Wu (2006) identified a QBO signal in gravity wave activity at equatorial latitudes using radio occultation temperature profiles. Ern et al. (2008) found QBO-related variations in gravity wave amplitudes as estimated from temperature data from the SABER satellite instrument as well as ECMWF temperatures. Ern et al. (2014) estimated gravity wave momentum fluxes based on HIRDLS and SABER temperature observations and found that gravity waves with intrinsic phase speeds less than 30 m s$^{-1}$ interact with the QBO.

Lindgren et al. (2020) estimated QBO-related variability in gravity wave amplitudes and spectral slopes in the 20°N-20°S domain using Loon balloon observations. They showed that gravity wave amplitudes tend to be higher during the westerly phase of the QBO than during the easterly phase and that spectral slopes are slightly steeper during the westerly phase for frequencies up to about 200 cycles day$^{-1}$.

The wave driving of the QBO has been explored in many prior studies. In contrast, the focus of the present study is to characterize the gravity waves observed during opposing QBO phases. The paper is organized as follows. In Section 2 we provide a brief description of the ERA5 data and the methodology. In Section 3 we contrast the gravity wave spectra observed during opposing phases of the QBO. We show that the two-sided zonal wavenumber-frequency power spectra of vertical velocity and the cospectra of the vertical flux of zonal momentum are discernibly different during westerly and easterly phases. By means of compositing analysis, we show that these differences are consistent with the distinctions between the three-dimensional structure of the waves as they disperse and propagate upward through background zonal wind profiles characteristic of easterly and westerly QBO phases. A brief discussion of the results and concluding remarks are presented in section 4.

## 2. Data and methodology

The data and methods used in this study closely follow those in Pahlavan et al. (2023), and are described only briefly here. We retrieve ERA5 hourly data at 0.25° × 0.25° horizontal resolution on standard pressure levels for the period 1979-2019. In order to investigate the gravity wave spectra in opposing phases of the QBO, two-sided zonal wavenumber ($k$)-frequency ($\omega$) spectra and cospectra are calculated for dynamical variables on the latitude circles between 10°S-10°N and averaged to obtain a single spectrum and cospectrum. All calculations are averaged over specified time intervals corresponding to westerly and easterly phases of the QBO.

We examine the power spectrum of vertical velocity, which emphasizes the role of higher frequency waves, and the cospectrum of vertical velocity and zonal wind; i.e., the vertical flux of zonal momentum, whose vertical derivative is proportional to the zonal force exerted on the mean flow by gravity waves. Including the cospectrum is informative, since the vertical velocity perturbations tends to emphasize vertically propagating waves, whereas the zonal wind spectrum emphasizes waves that propagate zonally.

The phase speed spectra of gravity waves with frequencies higher than 0.5 cpd are derived from the zonal wavenumber-frequency spectra. Each value in the $(k, \omega)$ domain is associated with a phase speed $c = \omega/k$. The phase speed spectrum is produced by grouping all of the $(k, \omega)$ pairs into 2 m s$^{-1}$ phase speed bins, and summing values in each phase speed bin. This binning procedure yields the gravity wave contribution to the momentum flux as a function of ground-based zonal phase speed.

In computing the spectra, the QBO phases are defined based on the monthly mean zonal wind at the 50 hPa level. Values above 2 m s$^{-1}$ are considered a westerly phase (QBO W), while those below -2 m s$^{-1}$ signify an easterly phase (QBO E). This partitioning yields 228 months of QBO W, and 204 months of QBO E. Composite QBO W and QBO E zonal mean zonal wind profiles are shown in Fig. 1a.

The gravity wave composites are created using the same methodology as in Pahlavan et al. (2023), except that the three-dimensional vertical velocity field does not need to be high-pass filtered, because it is already dominated by the high-frequency perturbations associated with gravity waves. The composites are based on the strongest downwelling events (i.e., those that fall in the top 10% of the frequency distribution) in a reference time series of tropospheric vertical velocity at individual grid points equally spaced along the equator at 1° intervals (i.e., 360 grid points). The composites for individual grid points are shifted in longitude so that they share a common reference grid point and then averaged together to form a single composite. The reference timeseries is the average of the standardized 700 and 300 hPa vertical velocities, which closely resembles the principal component of the leading mode of variability of the vertical profile of vertical velocity in the tropical troposphere. It is associated with the heating profile in deep convection. In the composites, QBO W and QBO E are defined on the basis of the state of the zonally averaged zonal wind averaged over 10°S-10°N at the 20 hPa level. Times when zonal wind is from the west and exceeds 2 m s$^{-1}$ are assigned to the QBO W composite and times when it is from the east and exceeds -16 m s$^{-1}$ are assigned to QBO E. This partitioning yields 156 months of QBO W, and 227 months of QBO E. The corresponding composite QBO W and QBO E zonal mean zonal wind profiles are shown in Fig. 1b.

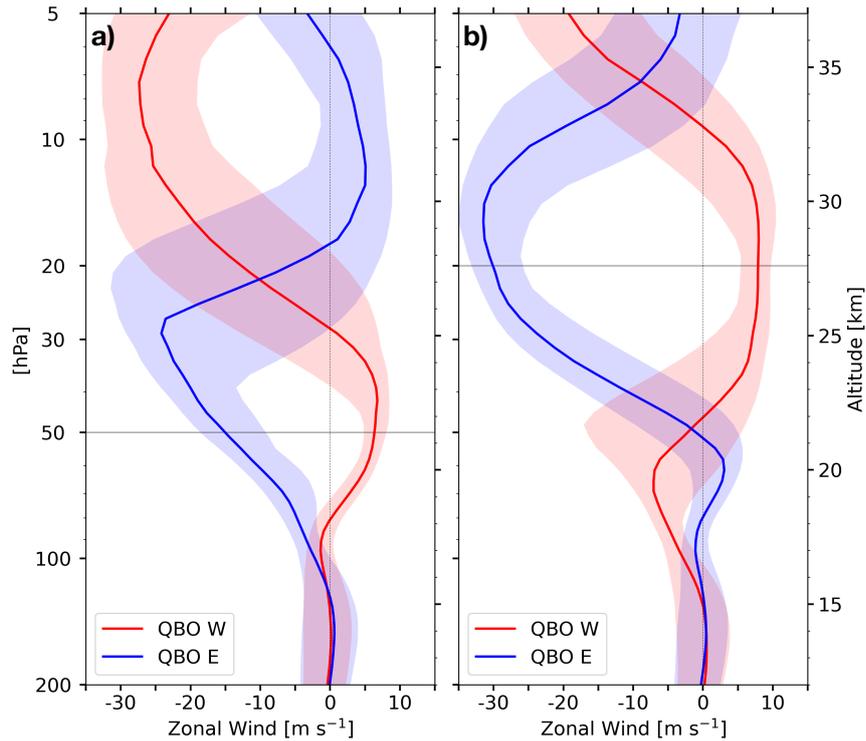

**Fig. 1.** Vertical profiles of zonally averaged zonal wind averaged over 10°S-10°N based on the QBO phases at the (a) 50 hPa and (b) 20 hPa levels for 1979-2019. The partitioning based on the 50 hPa and 20 hPa levels results in 228 and 156 months of the QBO westerly phase (QBO W) and 204 and 227 months of the QBO easterly phase (QBO E), respectively. Solid lines indicate the median, and shadings indicate the quartiles (25% and 75%) of the monthly data.

## 3. Results

Figure 2 shows time–height sections of the vertical flux of westerly momentum over a 10-year interval averaged over 10°S–10°N, superimposed on the zonal wind field. QBO-related wave-mean flow interactions are evident and consistent with theory: filtering of the wave spectrum by the mean flow systematically reduces the momentum flux of waves propagating in the same direction as the background flow. As a result, the net upward flux of zonal momentum is westward during the westerly phase of the QBO and eastward during the easterly phase. The contributions from high-frequency (i.e., $\omega \geq 0.5$ cpd), and low-frequency (i.e., $\omega < 0.5$ cpd) waves to the total flux, shown in middle and bottom panels of Fig. 2, respectively, exhibits several interesting differences: the westward momentum flux is dominated by high-frequency gravity waves; the fluxes by the low-frequency waves are largely confined to the lower stratosphere, below the 30 hPa level, while the fluxes by the high-frequency waves extend through the depth of the stratosphere.

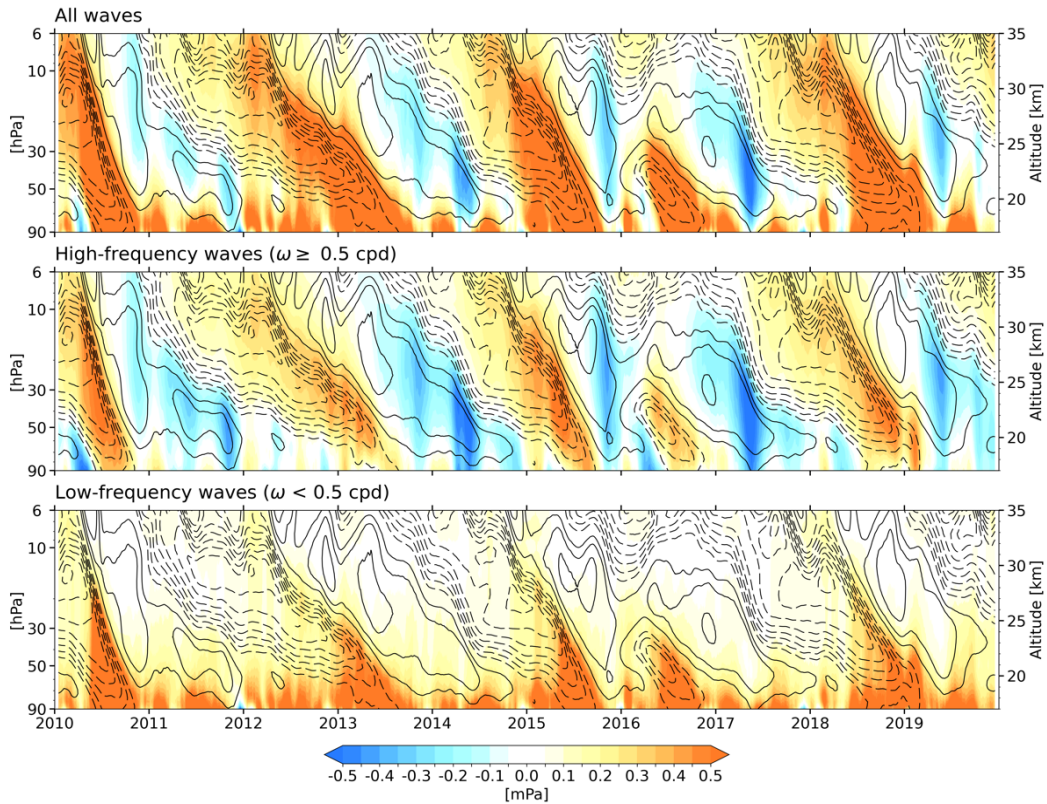

**Fig. 2.** Time–height sections of the vertical flux of zonal momentum (color shading) averaged over 10°S-10°N for (top) all waves, (middle) high-frequency waves (i.e., $\omega > 0.5$ cpd), and (bottom) low-frequency waves, which is obtained as the difference between the top and middle panels. Contours indicate the zonally averaged zonal wind. The contour interval is 5 m s$^{-1}$, westerlies are solid, easterlies are dashed, and the zero contour is omitted. The results are based on hourly data, smoothed by a 30-day running mean.

The low-frequency waves, which are dominated by Kelvin waves are most prominent at the ~90 hPa level. During the phase of the QBO when easterlies occupy the lower stratosphere, these waves disperse and propagate upward with relatively little dissipation until they encounter the first westerly shear zone, where their dissipation induces an eastward acceleration and the descent of the shear zone into the lower stratosphere. The opposite is observed during westerly phase of the QBO, except that the vertical flux of westward momentum by the gravity waves is more intermittent, as reflected in the more irregular descent rate of the easterly shear zones.

*3.1. Spectral analysis*

First let us consider the properties of power spectra of vertical velocity ($w$) during opposing phases of the QBO. Figure 3 shows the two-sided zonal wavenumber-frequency power spectra at various levels. Positive zonal wavenumbers correspond to eastward and

File generated with AMS Word template 2.0

negative values to westward propagating waves. Lines of constant slope corresponding to the gravity waves with phase speeds of 23 and 49 m s$^{-1}$ are plotted for reference. See Pahlavan et al. (2023) for further discussion of these spectra and their characteristics, and the slight preference for the two reference phase speeds, which are associated to the leading eigenmodes of convection in the tropical troposphere. The 100 hPa spectrum is representative of waves propagating upward from the troposphere into the stratosphere. The 50 and 20 hPa spectra illustrate the "wind-filtering" of the waves as they disperse and propagate upward through the lower stratosphere.

At the 100 hPa level, the *w* spectra for QBO W and E are fairly similar, and as a result, the differences between them, shown in the bottom right panel are quite small. QBO E exhibits slightly more power (i.e., variance) throughout nearly the entire spectrum. Considering that the background wind at the 100 hPa and below is virtually the same during QBO W and QBO E (see Fig. 1a), the higher variance during QBO E might be indicative of a higher rate of wave generation rather than stronger upward propagation.

The *w* spectra shown in Fig. 3 are indicative of a weakening of the waves propagating upward from the 100 to the 20 hPa level, which occurs irrespective of the phase of the QBO. This is to be expected, since the waves propagating upward through this layer encounter a wide range of background zonal wind profiles, as evidenced by the substantial width of the bands of interquartile shading in Fig. 1a. However, upon close inspection (noting that the amplitude scale is logarithmic), it is evident that the eastward propagating waves are damped

File generated with AMS Word template 2.0

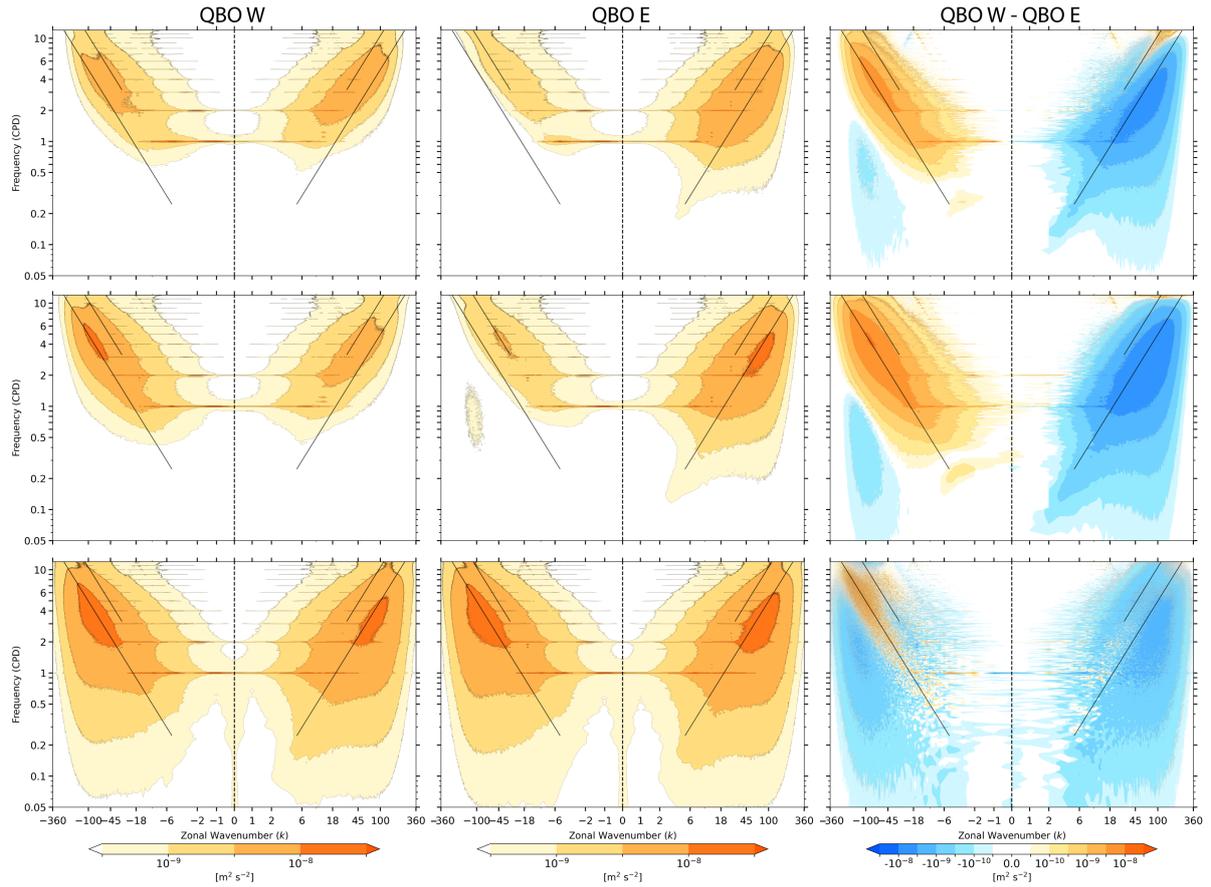

**Fig. 3.** Two-sided zonal wavenumber-frequency power spectra for vertical velocity at the (top) 20 hPa, (middle) 50 hPa, and (bottom) 100 hPa levels, based on the QBO phases. The spectra are shown on a logarithmic scale for wavenumber, frequency, and spectral density in unit of variance per unit area, from an average of the spectra computed at 1° latitude intervals between 10°S-10°N based on hourly data. The spectral power is multiplied by $k \times \omega$ to preserve the property that (power × area) be proportional to variance anywhere on the plot. The contour interval corresponds roughly to a factor of 3. The sloping straight black lines correspond to phase speeds of ±49 m s$^{-1}$ (upper) and ±23 m s$^{-1}$ (lower). The scale of zonal wavenumber is linear from $k$ = -1 to 1.

substantially more during QBO W and vice versa. Such "wind-filtering" is responsible for the pronounced dipoles (i.e., the contrasting colors) in the difference spectra in the top and middle panels in the right column of Fig. 3. The differences between the QBO E and QBO W spectra are consistent with Lindzen and Holton's mechanistic model of the QBO, which is based on the principle that as a wave approaches its critical level, where its ground-based phase speed is equal to the background wind speed, its vertical wavelength and group velocity become small and it becomes more susceptible to dissipation. Consequently, in the QBO E the westward propagating waves are preferentially weakened due to the presence of the easterly background flow and vice versa. The wave dissipation, in turn, forces the background flow, leading to the downward propagation of alternating westerly and easterly wind regimes.

Figure 4 shows the phase speed spectra of the high-frequency waves (i.e., $\omega > 0.5$ cpd) during opposing phases of the QBO, which are derived from $w$ spectra in the manner explained in Section 2. At the 100 hPa level the variance is higher in QBO E, especially for eastward propagating waves, consistent with Fig. 3. Proceeding upward from the 100 to the 50 hPa level, we see that eastward propagating waves are more strongly dissipated during QBO W and westward propagating waves are more strongly dissipated during QBO E. As in Fig. 3, the dissipation is somewhat stronger for the westward propagating waves. It is also evident from Fig. 4 that the waves with slow phase speeds are significantly damped while propagating upward from the 100 to the 50 hPa level, shifting the phase speed peak toward higher values. At the 20 hPa level, the power of westward propagating waves approaches zero in the QBO E composite for waves with $|c| < 20$ m s$^{-1}$. The eastward propagating waves that reach the 20 hPa level, contribute to the forcing of the westerly phase of the semi-annual oscillation (SAO).

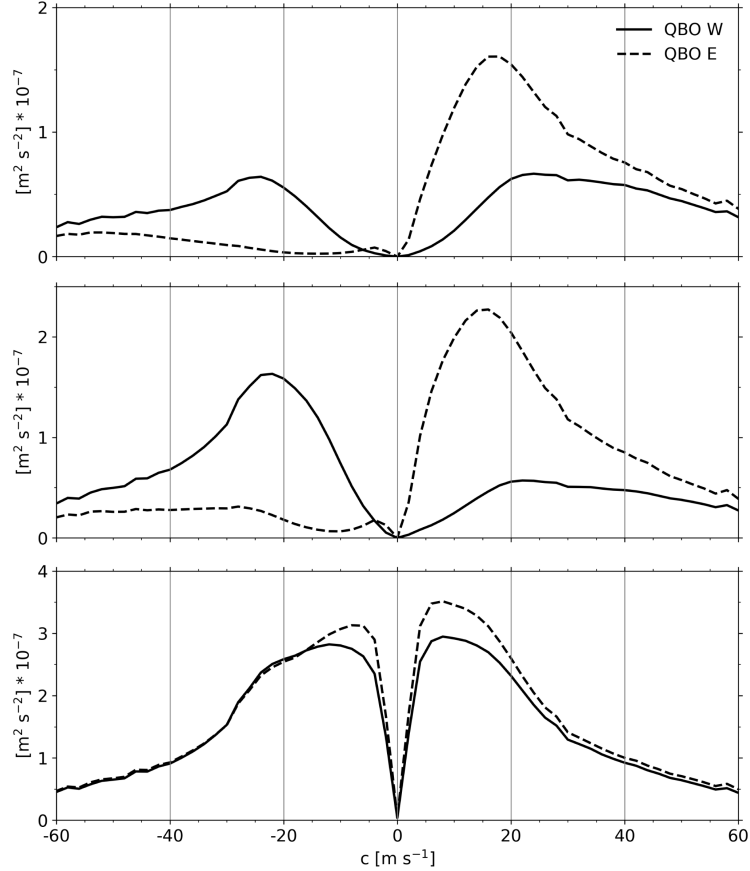

**Fig. 4.** Phase speed spectra of high-frequency waves (i.e., $\omega > 0.5$) obtained from vertical velocity power spectra at the (top) 20 hPa, (middle) 50 hPa, and (bottom) 100 hPa levels. Note that the scaling of the vertical axis is different at each level.

Now let us consider the spectra of vertical flux of zonal momentum ($F_M$) in opposing phases of the QBO. $F_M$ is the primary driver of the QBO, as shown in Figs. 8 and 9 of Pahlavan et al. (2021b), and its absolute value is a measure of the upward flux of wave activity. In the two-sided zonal wavenumber-frequency spectra of $F_M$, shown in Fig. 5, positive values are indicative of an upward flux of eastward momentum and vice versa.

In contrast to the power spectrum of $w$ shown in Fig. 3, which is dominated by higher frequency/smaller-scale gravity waves, low-frequency, planetary-scale waves and notably Kelvin waves make an appreciable contribution to the $F_M$ spectrum. Consequently, the $F_M$ spectrum is not as symmetric with respect to the direction of propagation of the waves as the $w$ spectrum.

At the 100 hPa level, close to the sources of the gravity waves, the momentum flux is strong and quite similar in the QBO W and E composites, so the differences are generally small. However, within a band of eastward propagating waves with zonal wavenumbers between 2 and 10 and frequencies < 0.2 cpd, i.e., in the range of Kelvin waves, the flux is substantially stronger during QBO E than during QBO W. Results of Abhik et al. (2019), and Sakaeda et al. (2020) suggest that convectively coupled Kelvin waves are in fact slightly stronger during QBO E. It should be noted that we have not considered the state of ENSO in our analysis. However, there is no sampling bias of QBO W/E composites toward its warm or cold polarity during this 41year-long record.

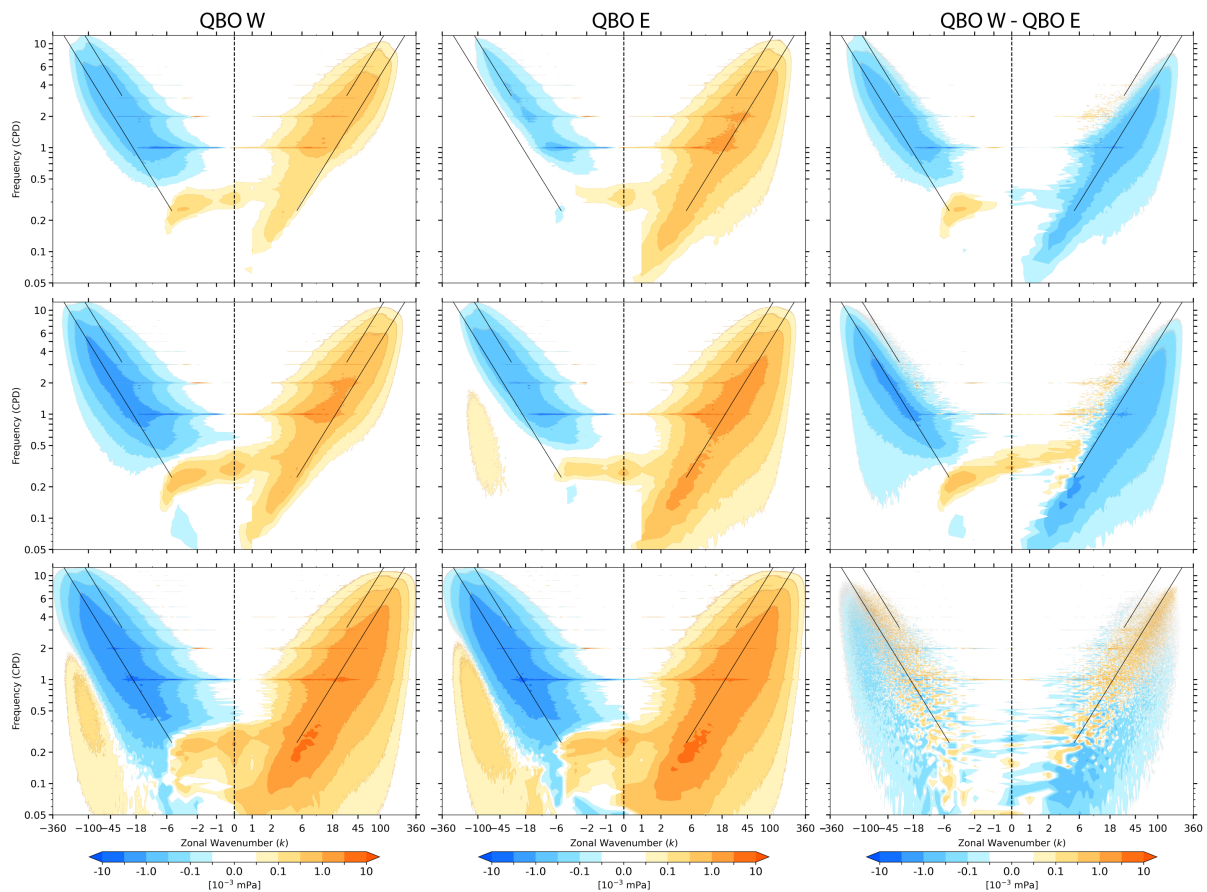

**Fig. 5.** Same as in Fig. 3, but for the vertical flux of zonal momentum.

As in the interpretation of the $w$ spectra in Fig. 3, the general decrease of $F_M$ from the 100 to the 20 hPa level is indicative of the existence of wave dissipation throughout the stratosphere. Consistent with the $w$ spectra, there is stronger dissipation of eastward propagating waves in QBO W, and stronger dissipation of westward propagating waves in QBO E, as reflected in the dominance of blue shading in the difference spectra. Waves with phase speeds that fall within the range of zonal wind speeds in the QBO (-35 to +20 m s$^{-1}$) are much more strongly attenuated than waves with higher phase speeds. Hence, proceeding upward from the 100 to the 20 hPa level, the peaks in the $F_M$ spectra are shifted toward higher phase speeds. The $F_M$ spectra exhibit a distinctive MRG wave signature: consistent with theoretical expectations, it is much stronger in QBO W than in QBO E at the 50 and 20 hPa levels.

Figure 6 shows the phase speed spectra of $F_M$ for gravity waves with frequencies higher than 0.5 cpd. The scales on the abscissas have been adjusted to make the weak $F_M$ at the higher levels more clearly visible. As in the wavenumber-frequency spectra shown in Fig. 5, the stronger damping of the more slowly propagating waves that interact with the QBO is reflected in the increasing prominence of higher phase speeds at the higher levels. Other notable features that is evident in both Figs. 4 and 6 at the 50 and 20 hPa levels are the peak in eastward propagating waves with phase speeds ~ 15 m s$^{-1}$ during QBO E and the peak in westward propagating waves with phase speeds ~ -25 m s$^{-1}$ during QBO W. These phase speeds fall within the range of convectively coupled Kelvin and MRG waves, and might be related to these two modes of variability in the tropical troposphere.

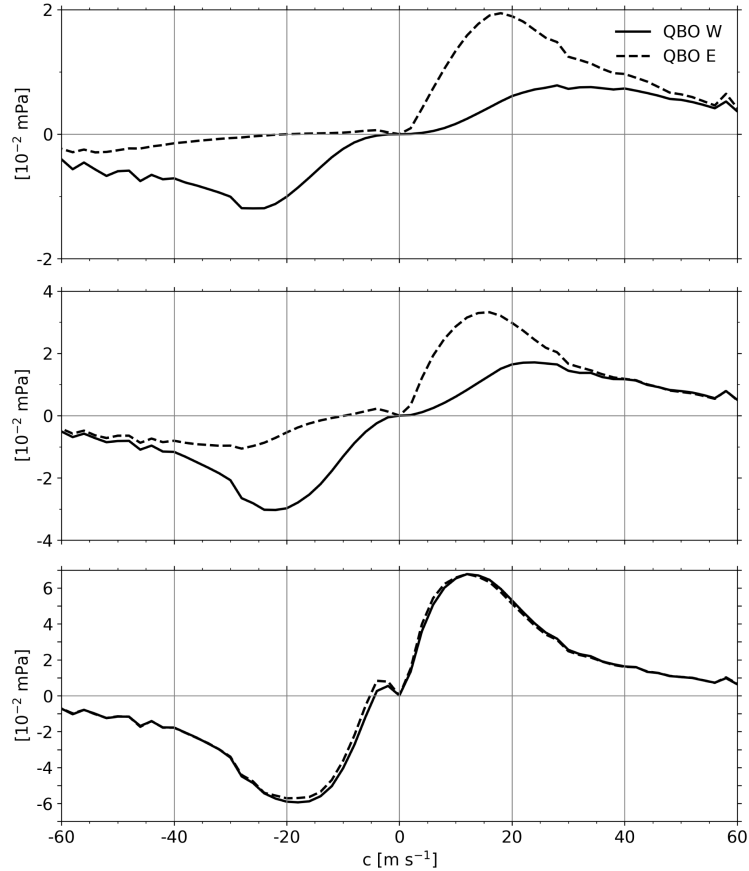

**Fig. 6.** Same as in Fig. 4, but for the vertical flux of zonal momentum. Note that the scaling of vertical axis is different at each level.

*3.2. Compositing analysis*

In Pahlavan et al. (2023), we examined the three-dimensional structure of gravity waves by compositing different variables based on time series of tropospheric vertical velocity at individual reference grid points. In this subsection, we extend that analysis to investigate the structure of waves propagating upward during opposing phases of the QBO, using the methodology described in Section 2. Note that in this subsection, QBO W and QBO E are defined on the basis of the sign of the zonal wind at the 20 hPa level rather than at the 50 hPa level (see Fig. 1b).

The top panels of Fig. 7 show composite horizontal maps of vertical velocity at the 10 hPa level in QBO W and E and the bottom panels show the corresponding equatorial longitude-height sections. The wind-filtering of the waves in the stratosphere is responsible for the east-west asymmetry of the wave structures. The waves are amplified and compressed upstream of the reference point and attenuated and stretched downstream of it. The asymmetries are more

File generated with AMS Word template 2.0

pronounced during QBO E (right panels) due to greater strength of the easterly regimes. In the longitude-height sections the waves tend to be dissipated (i.e., attenuated) as they disperse and propagate upward through the layer of strong zonal wind shear below the peak winds at the 20 hPa level: eastward propagating waves are dissipated during QBO W and westward propagating waves during QBO E. These results are consistent with previous observational and modeling studies (e.g., Yue et al. 2009; Vadas et al. 2012; Xu et al. 2015; Nyassor et al. 2021), which have reported that in the presence of weak background wind speeds, gravity waves propagate freely into the upper atmosphere so that their phase fronts are concentric circles, whereas when $|u| > 30$ m s$^{-1}$, they propagate mainly upstream and their phase fronts are elliptical, semi-elliptical, semi-circular, or arc-like.

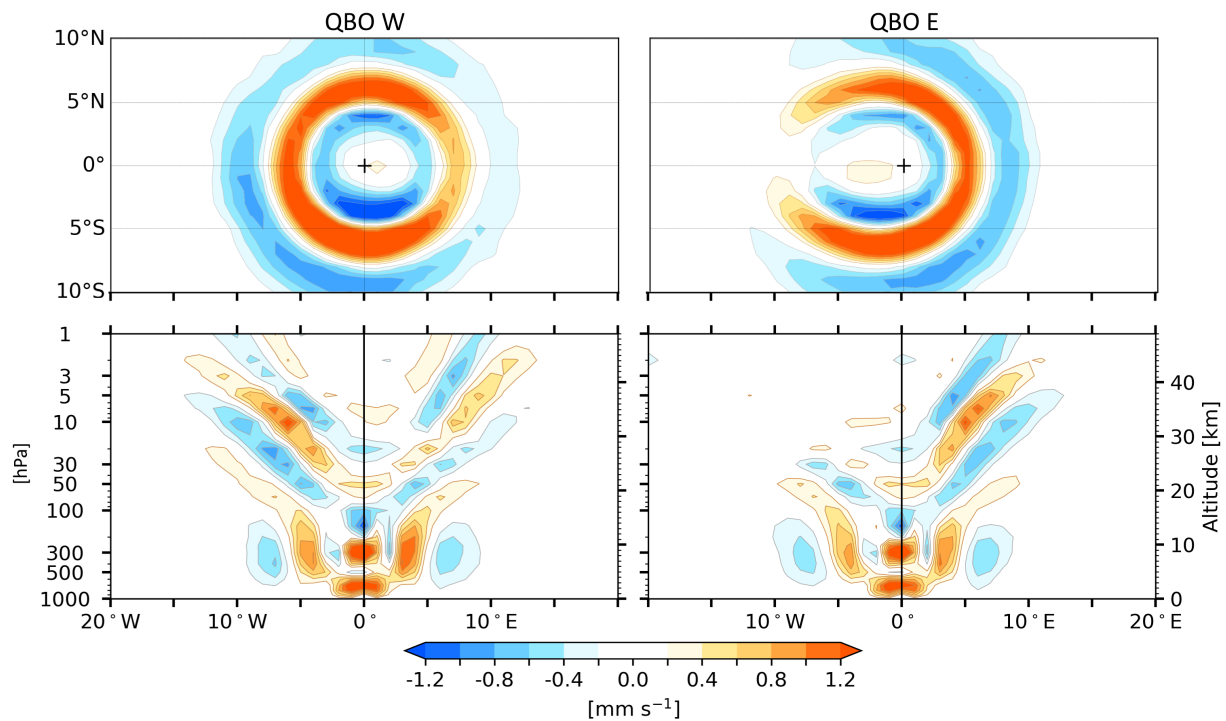

**Fig. 7.** Composites of high-pass filtered vertical velocity constructed based on the top 10% of strongest downwelling events at individual reference grid points along the equator, shown 8 hours after the events, using hourly data for 1979-2019, and based on the QBO phases at the 20 hPa level. The reference timeseries is the sum of the standardized 700 and 300 hPa vertical velocity [see Pahlavan et al. (2023) for more detail]. (top) Horizontal cross sections of vertical velocity at the 10 hPa level. (bottom) The corresponding longitude-height cross sections along the equator. The reference grid point is shown by a cross in the top panels.

It also should be noted that these composites are based on the deep tropospheric mode, defined by the sum of standardized 700 and 300 hPa vertical velocities (Pahlavan et al. 2023), for which the horizonal phase speed is ~50 m s$^{-1}$. Therefore, the waves can propagate through the weak westerly phase of the QBO without much dissipation, as can be seen in the left panels of Fig. 7.

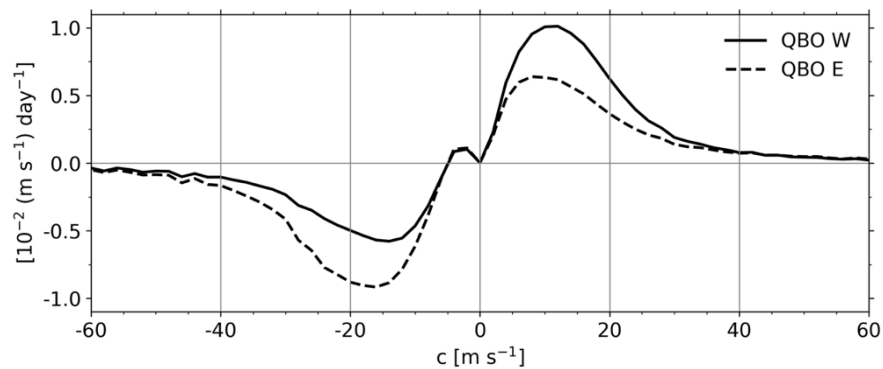

**Fig. 8.** Phase speed spectra of wave forcing due to high-frequency waves (i.e., $\omega > 0.5$), calculated as the divergence of vertical flux of zonal momentum between 100 and 50 hPa levels (i.e., the middle and bottom panels of Fig. 5).

## 4. Discussion and concluding remarks

Our results concerning the wind-filtering of gravity waves as they disperse and propagate upward through the lower stratosphere are consistent with the mechanistic model of the QBO in Lindzen and Holton (1968), which involves the interactions between vertically propagating gravity waves and the zonally averaged zonal flow in the absence of rotation. It is notable that the Abstract of their paper refers to "the interaction of long-period, vertically propagating gravity waves with the zonal wind" and it is clear from Section 2 of their text that the "long-period, vertically propagating gravity waves" connoted planetary-scale Kelvin and MRG waves. Yet the waves in their simple model were more like the small-scale, high-frequency inertio-gravity waves examined in our study. Their mention of Kelvin and MRG waves stems from the observational evidence available for these two modes of waves at that point in time (Maruyama 1967; Wallace and Kousky 1968). We have shown that both planetary waves and small-scale gravity waves participate in QBO-related mean flow interactions, but only the high-frequency gravity waves are capable of inducing westward as well as eastward accelerations of the mean flow, as required for explaining the oscillatory behavior of the QBO. Below the 50 hPa level, the dissipation of Kelvin waves makes a substantial contribution to the downward propagation of westerly wind regimes, but it is not crucial for explaining the existence of the QBO.

We have also considered how the QBO modulates the properties of the waves at 100 hPa, below the layer in which wind-filtering plays an important role. In agreement with prior studies of Knaff (1992), Collimore et al. (1998), and Sweeney et al. (2023), which reported evidence of more vigorous convection and cloud activity during QBO E than during QBO W, we show in Figs. 3 and 4 that the variance of the vertical velocity field is also somewhat higher during QBO E. It is notable that almost all of the enhanced variance is associated with waves with phase speeds less than 20 m s$^{-1}$. Another unexpected finding is that at the 100 hPa level, the upward fluxes of zonal momentum in QBO W and QBO E are virtually identical at all phase speeds for waves with frequencies higher than 0.5 cpd (Fig. 6).

The wavenumber-frequency spectra presented in Figs. 3 and 5 show that eastward propagating waves experience greater damping during QBO W, and that westward propagating waves are damped substantially more during QBO E, both of which align with expectations. Yet, a striking observation from these figures is that as the waves propagate upward from 100 to 20 hPa level, both eastward and westward propagating waves undergo strong dissipation, regardless of the phase of the QBO. This finding implies that in ERA5, wave dissipation occurs ubiquitously, without sufficient selectivity. This is better shown with phase speed spectra of wave forcing (or wave dissipation) due to high-frequency waves, derived from the divergence of the vertical flux of zonal momentum between 100 and 50 hPa levels, shown in Fig. 8. Eastward propagating waves during QBO E and westward propagating waves during QBO W undergo an equivalent level of background dissipation [$7.5 \times 10^{-2}$ (m s$^{-1}$) day$^{-1}$], cancelling out 64% of the westward forcing during QBO E and 70% of the eastward forcing during QBO W. This implies that if the background dissipation could be significantly reduced, the forces driving QBO descent might be significantly stronger.

These results are consistent with the findings of Holt et al. (2016), who used 2-yr global simulation with 7-km-horizontal-resolution of the Goddard Earth Observing System Model version 5 (GEOS-5) to study the wave forcing of the QBO. Their study also determined that due to background dissipation, the EP-flux divergence is reduced by about half during QBO E and reduced by 84%–95% between 50 and 10 hPa during QBO W. This unrealistic damping is most likely due to the degree of explicit divergence damping and implicit dissipation associated with the numerical scheme. This factor might be a prime contributor to the poor representation of the QBO in the lower stratosphere by most of the models.

Furthermore, in a recent study, Bramberger et al. (2023) used temperature observations from Strateole-2 balloon flights to study tropical waves. Their findings highlighted that

File generated with AMS Word template 2.0

higher-frequency wave packets with periods shorter than one day are underestimated in ERA5 reanalysis. While their study focused on one event, our Figs. 3, 5, and 8 suggest that pronounced and unrealistic wave damping is a pervasive feature in ERA5. In our earlier study (Pahlavan et al. 2021a), we showed that in ERA5, the resolved wave forcing accounts for at least half the required wave forcing for the downward phase propagation of the QBO. Our findings in this study suggest that, should ERA5 properly manage wave dissipation, this contribution could be considerably more substantial.

*Acknowledgments.*

This research is supported by the NSF Grant AGS-2202812. The authors declare no conflicts of interest.

*Data Availability Statement.*

ERA5 model-level data were downloaded from ECMWF's MARS archive, while pressure-level data were downloaded from the Copernicus Climate Data Store (https://cds.climate.copernicus.eu/).